# First Light SOFIA Instruments[1]

Alfred Krabbe[a], Sean C. Casey[b]

[a] Physics Department, UCB, Berkeley
[b] Universities Space Research Association

## ABSTRACT

The Stratospheric Observatory For Infrared Astronomy SOFIA[2] will become operational with the next two years. It will be the biggest astronomical airborne observatory ever build, comprising a 3m-class telescope onboard a Boeing 747SP. A suite of first-light instruments is under development striving for cutting edge technology to make SOFIA a milestone in infrared astronomy. Here we present an overview over the instrumentation and an update on the current status.

## 1. SOFIA

SOFIA, a bi-national American-German project, is opening a new era in MIR/FIR astronomy. From 2004 on, SOFIA will offer regular access to the entire MIR and FIR wavelength range between 5 micron and 300 micron part of which is otherwise inaccessible from ground. Sofia's 2.7m-size mirror together with it's optimized telescope system combines the highest available spatial resolution with excellent sensitivity. SOFIA can operate in both celestial hemispheres, will always fly the latest instrument technology, and will be available for the next two decades to come. With the aircraft currently undergoing major modification, the telescope being delivered this summer and the SOFIA Science and Mission Operation Center (SSMOC) close to completion, SOFIA as an observatory is becoming reality. For a more detailed overview, the reader is referred to proceedings of SPIE vol. 4014 (2000) and 4847 (2002), a large fraction of which is devoted to SOFIA.

## 1. INSTRUMENTATION

Parallel to the development of the observatory, USRA issued a call for proposals for NASA that led to the selection in 1997 of seven US instrument teams. These teams are currently developing and building state-of-the-art cutting edge technology instruments to make SOFIA a milestone in infrared astronomy. A similar call, issued by the German Aerospace center (DLR), led to the selection of two instruments. Criteria for the selection were instruments that would build from the strength of SOFIA and address the best science SOFIA could possibly deliver. Table 1 highlights the range of scientific topics to be addressed by Sofia's first light instrumentation.

**Table 1**

| High Angular Resolution | High Spectral Resolution |
|---|---|
| • **Imaging studies** | • **Spectroscopic studies** |
| - **Protostellar environments** | - **Accretion and outflows** |
| - **Young star clusters** | - **Gas excitations** |
| - **Molecular clouds** | - **Gas dynamics** |
| - **Galaxies** | |
| • **Multi-color studies** | |
| - **Dust temperature** | |
| - **Optical depths** | |
| - **Dust compositions** | |
| - **Ionization source** | |

---

[1] Invited talk presented at SPIE 4818, *Infrared Spaceborne remote Sensing X*, Seattle, July 2002

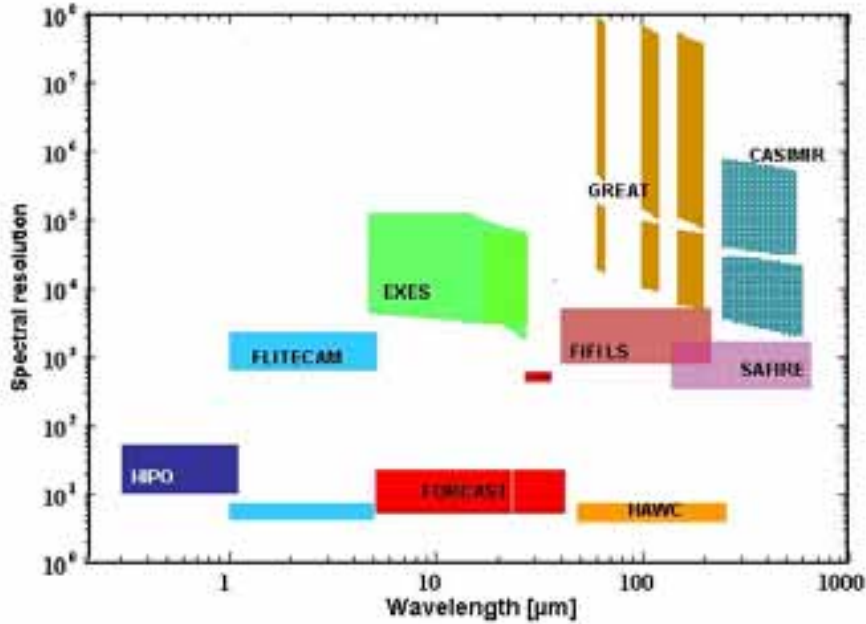

**Figure 1** Spectral resolution and wavelengths coverage of SOFIA's first light instrumentation.

The SOFIA science instruments of the first generation come in three types:

- **Facility instruments**
  Facility instruments will be operated by USRA[3] for the scientific community. These will be the main workhorses of the observatory. These instruments are all developed by American teams:

    - FLITECAM  Near-IR test camera 1 – 5 micron         PI I. McLean, Univ. Calif. Los Angeles
    - FORCAST   Mid-IR camera 5 – 40 micron              PI T. Herter, Cornell Univ.
    - HAWC      Far-IR camera 40 – 300 micron            PI A. Harper, Univ. of Chicago

- **PI instruments**
  These instruments are operated on a shared risk basis via the PI teams in a collaborative mode with the observing scientist. Instruments in this category are

    - CASIMIR   Heterodyne Spectrometer 250 – 600 micron    PI J. Zmuidzinas, CalTech
    - EXES      Echelon Spectrometer 5 – 28 micron          PI J. Lacy, Univ. of Texas
    - GREAT     Heterodyne Spectrometer 6 – 250 micron      PI R. Güsten, MPIfR, Germany
    - FIFI-LS   Field Imaging Spectrometer 40 – 210 micron  PI A. Poglitsch, MPE, Germany
    - SAFIRE    Fabry-Perot-Spectrometer 145 – 655 micron   PI H. Moseley, GSFC

  These instruments are mostly spectrometers in the range from 5 micron to 600 micron. GREAT and FIFI-LS are the instruments provided by German institutions. SAFIRE may not be available at first light but a year later.

- **Special instruments**
  These are also PI instruments but they serve a special purpose and thus are addressing special needs.
    - HIPO      Visible Occultation CCD 0.35 – 1.1 microns  PI E. Dunham, Lowell Observ.

---

[2] http://www.dlr.de/SOFIA, or http://www.sofia.arc.nasa.gov/
[3] Universities Space Research Association is the NASA contractor operating SOFIA for the first 5 years.

The instruments and their current status will be describes in more detail in the following sections. Figure 1 summarizes the wavelengths coverage and spectral resolution range of the different first-light instruments.

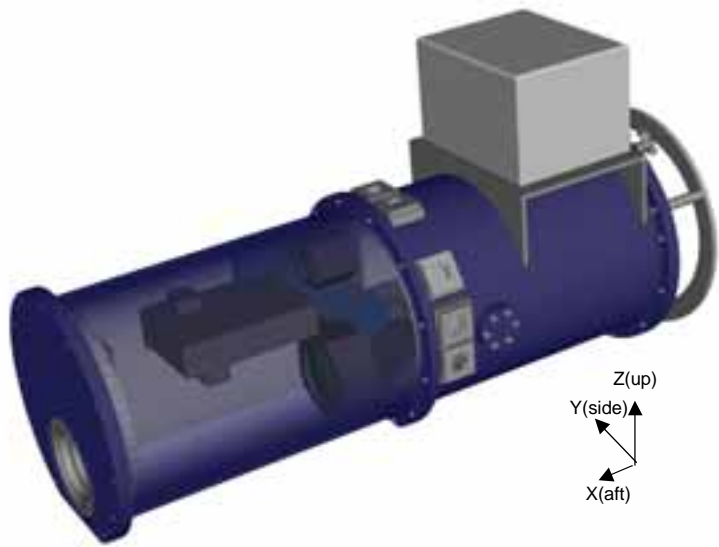

**Figure 2** Sketch of the FLITE-CAM experiment. Light enters through the window on the left.

## 1. FACILITY INSTRUMENTS

**FLITECAM**

FLITECAM (Figure 2) is being developed at he UCLA Infrared Imaging Detector Laboratory. Its prime purpose is to test the SOFIA telescope imaging quality from 1.0 to 5.5 microns, using a 1024 × 1024 InSb ALADDIN II array. Once the telescope test flights are finished, FLITECAM will be available to the science community. FLITECAM's field of view of 8 arcmin in diameter, with a plate scale of 0.47 arcsec per pixel, is one of the largest of any facility camera. Grims are also available to provide moderate resolution spectroscopy of R ≈ 1000 – 2000, depending on the slit width, with direct ruled ZnSe grisms. The detector readout electronics will be able to operate the detector array at all its planned

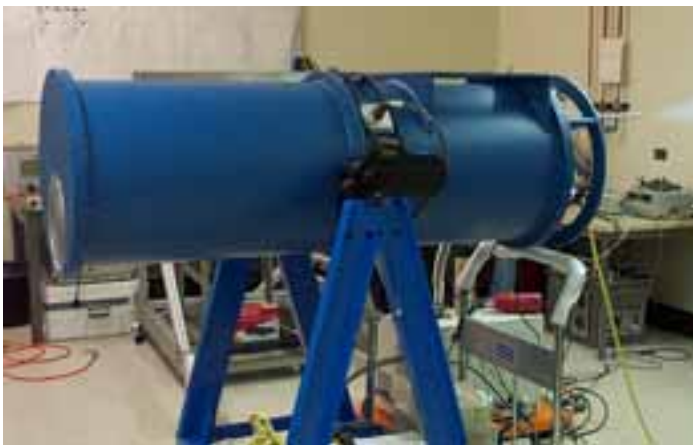
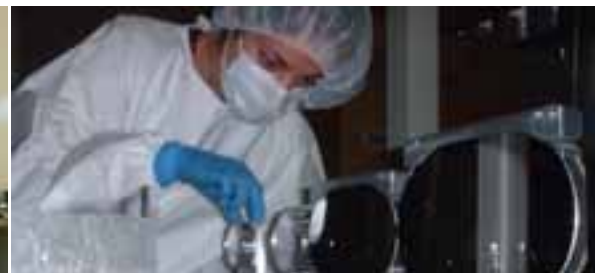

**Figure 3** (left) Testing the FLITECAM dewar.

**Figure 4** (above) Assembling the FLITECAM optics.

observation modes, including occultations, pupil viewing, telescope-nodding, high-speed shift-and-add, and optionally chopping at the longer wavelengths. During the test phase, FLITECAM will often be jointly operating with the high-speed occultation photometer HIPO (see below), to run image quality tests simultaneously at visible and IR wavelengths.

The dewar has been delivered and is now being tested (Figure 3). The optics has been assembled and is being aligned (Figure 4). Most of the other mechanics are ready and are being integrated. The electronics is all in hand: Data acquisition system, mechanism control, temperature control, pressure monitor, LHe monitor, and power controller. Next major steps will be integration and testing. First light at a ground-based telescope is being scheduled for fall 2002.

**FORCAST**

The fain object IR camera for SOFIA FORCAST is a two-channel camera (Figure 5) with selectable filters for continuum imaging in the 5 – 8 micron, 17 – 25 micron, and/or 25 – 40 micron regions. The design supports simultaneous imaging in the two channels. Using the latest 256 × 256 Si:As and SiSb blocked-impurity-band (BIB) detector array technology to provide high-sensitivity wide-field imaging, FORCAST will sample images at 0.75 arcsec/pixel and have a 3.2 arcmin × 3.2 arcmin instantaneous field of view. Imaging is diffraction limited for wavelengths larger than 15 microns. Since FORCAST operates in the wavelengths range were the seeing from SOFIA is best (Davidson 2000), it will provide the highest possible spatial resolution possible with SOFIA. FORCAST will eventually support a spectroscopic mode using silicon grisms, mounted in the filter wheels. The achievable spectral resolution at 5 – 8 micron and 17 – 40 microns will be R ≈ 300 and R ≈ 1000.

Each long wavelengths band-pass filter ($\lambda > 25$ microns) consists of a double half-wave interferometer (DHWI)[2,3] and a blocking filter. The DHWI is a resonant cavity formed with tree mirrors instead of two in the conventional Fabry-Perot interferometer (FPI). The additional interfering surface results in a bandpass profile that is much sharper than the Lorentzian profiles of FPIs, and therefore has much better out-of-band rejection than is possible with an FPI. Figure 6 shows the DHWI components ready to be tested. Figure 7 shows some of the optical components ready to be integrated. At this time the cryostat is being fabricated and delivery is expected in early fall this year. All Detectors are in house and are being tested. The cold electronics has been designed and the warm electronics has been delivered by Wallace Instruments. The integration of the optical bench is expected to be completed by February 03.

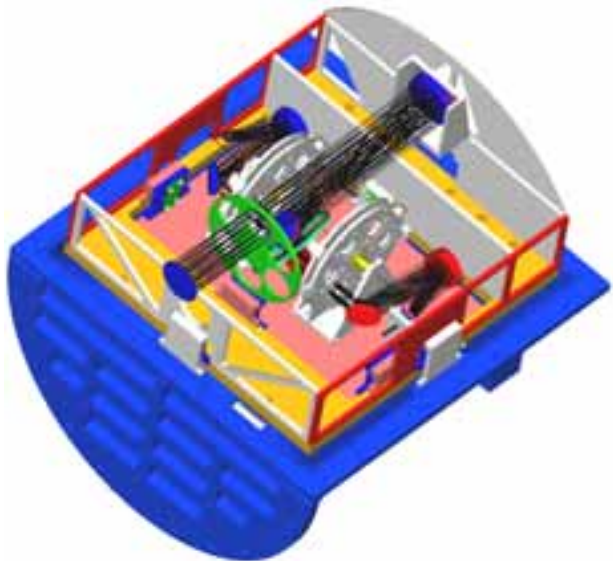
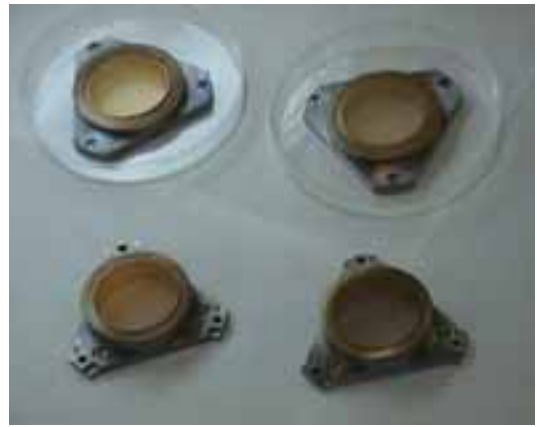

**Figure 5** (left)  Sketch of the optical dual channel set-up of FORCAST.
**Figure 6** (right) Fabry-Perot and double half-wave interferometer DHWI components ready for testing

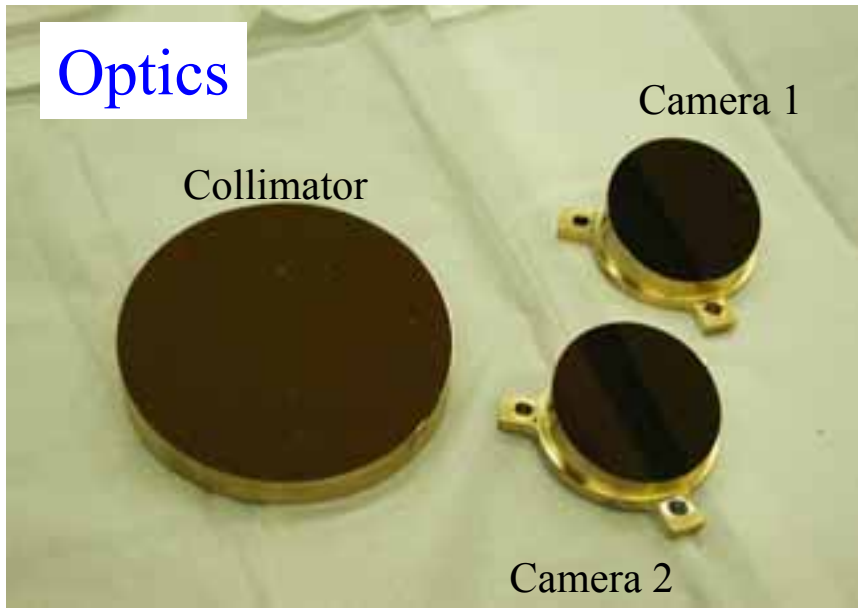

**Figure 7** FORCAST optics

## HAWC

HAWC is a far-IR camera designed to cover the 40 – 300 micron spectral range at the highest possible angular resolution (Figure 8). Four selectable broadband filters peak at 58, 90, 155, and 215 microns and provide a spectral width of $\Delta\lambda/\lambda$ 0.23, 0.1, 0.2, and 0.23 resp.. HAWC will utilize a 12 × 32 array of bolometer detectors constructed using the ion-implanted pop-up detector technology being developed at Goddard Space Flight Center. This new technology enables construction of closely packed, two-dimensional arrays of bolometers with high quantum efficiency and area filling factors of greater than 95% (Figure 11). The array will be cooled by an adiabatic demagnetization refrigerator (Figure 10) and operated at a temperature of 0.2 K.

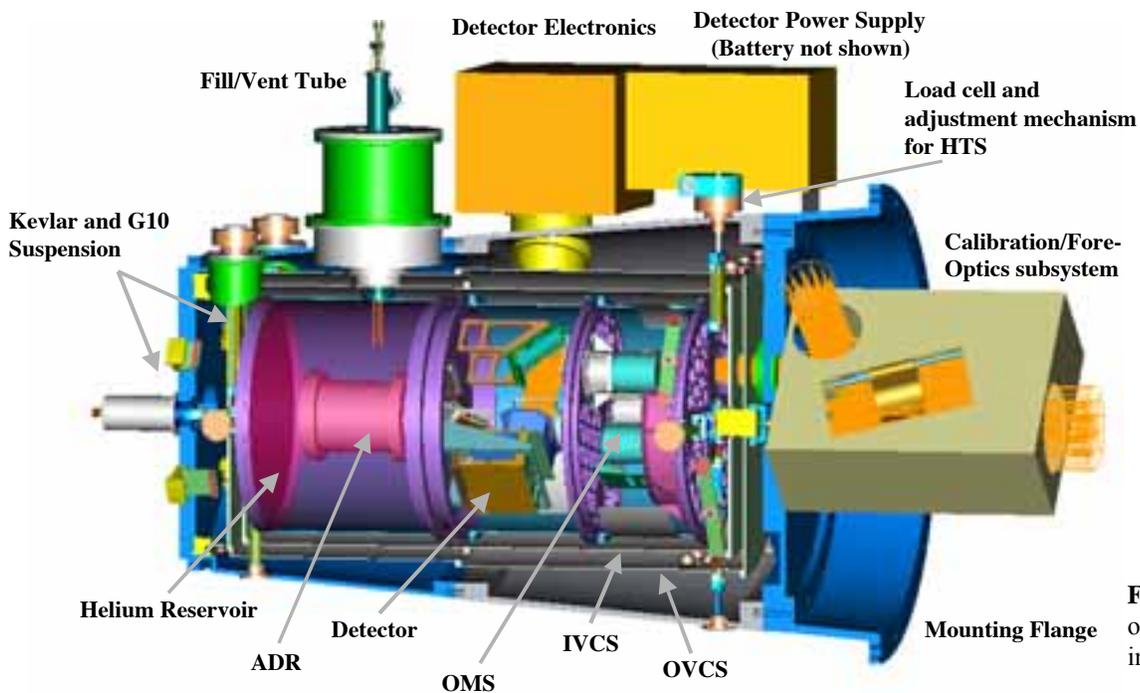

**Figure 8** Sketch of the HAWC instrument.

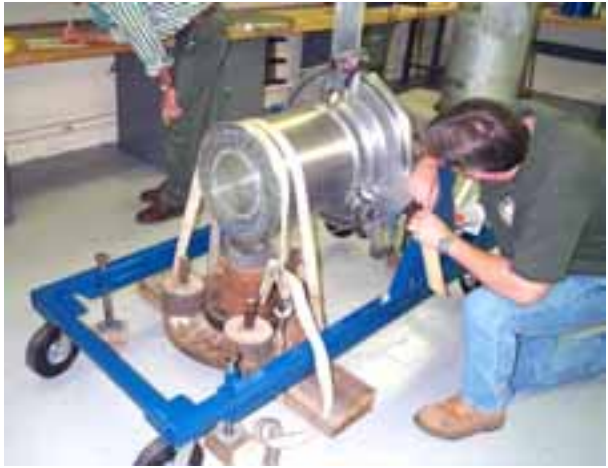 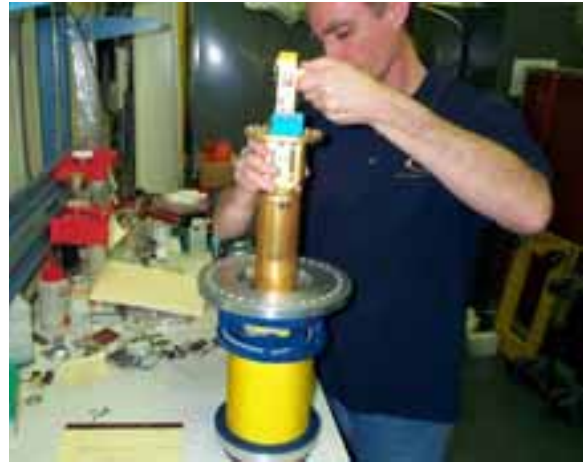

**Figure 9**  HAWC test cryostat suspension load test as required by the Federal Aviation Administration (FAA).

**Figure 10**  Assembly of the adiabatic demagnetization refrigerator.

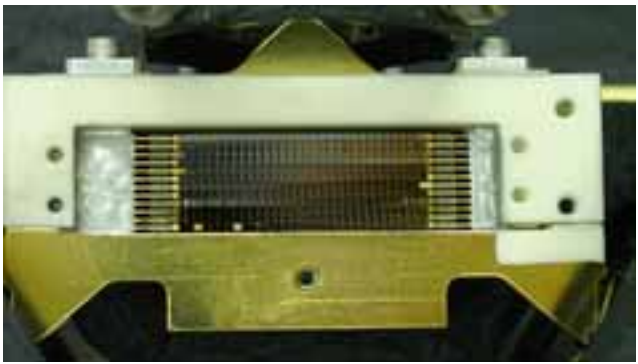

**Figure 11**  (left)  12 × 32 test array of pop-up bolometers developed by NASA GSFC.

**Figure 12**  (below)  First astronomical image with the test array was obtained at CSO on Jupiter at 350 micron. 2 out of 3 JFET drawers were installed (i.e. 8 rows functional).

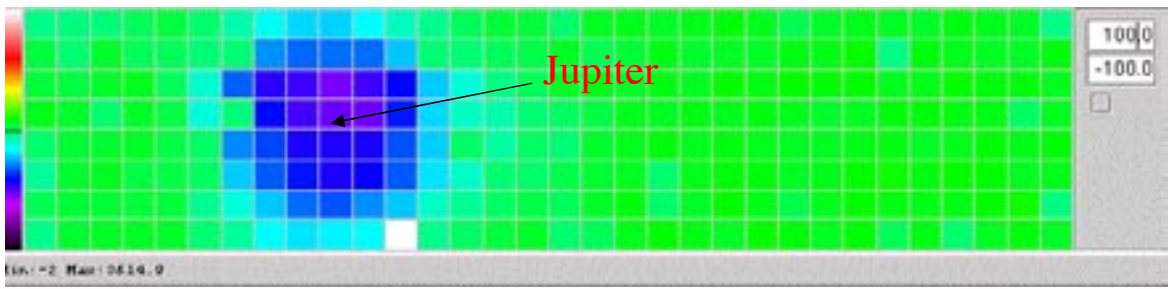

Meanwhile, the test dewar has been delivered (Figure 9). The first batch of detectors have also been delivered and mounted in another test dewar (Figure 11) for observations at CSO. A first image of Jupiter has been obtained at CSO by the team at a wavelength of 350 microns with the camera SHARC II demonstrating that these bolometer arrays are operational and providing the expected level of performance. The optics for HAWC is in the process of fabrication. The design of the HAWC flight dewar is being finalized for airworthiness certification.

## 2. PI INSTRUMENTS

**GREAT**

GREAT is a modular dual-channel heterodyne spectrometer offering a spectral resolution of $10^{6-7}$. At first light, three scientifically selected frequency windows will be available:
- A low-frequency band, 1.6 – 1.9 THz (157 – 189 micron), covering – among other lines – the important atomic fine-structure transition of ionized carbon.
- A mid-frequency band at 2.6 THz (115 micron), centered on the 1-0 transition of deuterated molecular hydrogen (HD) and the rotational ground-state transition of OH($^2\Pi_{3/2}$).
- A high-frequency channel at 4.8 THz (63 micron) centered at the fine-structure transition of atomic oxygen.

The instrument will consist of a front-end with 2 independent cryogenic dewars (Figure 13), mounted to the telescope flange via a common "optics box" with the local oscillator, diplexer, the polarization beam splitter, the calibration unit, and (optionally) a single sideband filter. The mixers are diffusion-cooled Nb or lattice-cooled NbN hot electron bolometers, depending on performance (Figure 14). The backend will include an initial choice of an acousto-optical array spectrometer with 4 × 1 GHz wide bands of 1 MHz spectral resolutions and an ultrahigh-resolution chirp transform spectrometer with 180 MHz bandwidth and 45 kHz spectral resolution.

Traditionally, local oscillators (LO) have been optically pumped far-IR ring laser. However, this type of LO is not continuously tunable and the necessity of finding a sufficiently strong laser line near the desired observing frequency poses a potential limit for the usage of such systems. For the SOFIA instrumentation, far-IR laser are the baseline option for the high frequency bands and the fallback option for the lower frequencies. In contrast, the development of tunable "high"-power solid state LOs is rapidly advancing into the THz regime, heavily stimulated by the HIFI instrument on the Herschel (former FIRST) satellite[4] (Figure 15). They are based on GaAs p-HEMT MMIC power amplifiers, operating in the range of 75 – 100 GHz that feed a chain of sub-millimeter-wave frequency multipliers. These solid state LOs have now become the baseline for the lower frequency bands (see also CASIMIR below).

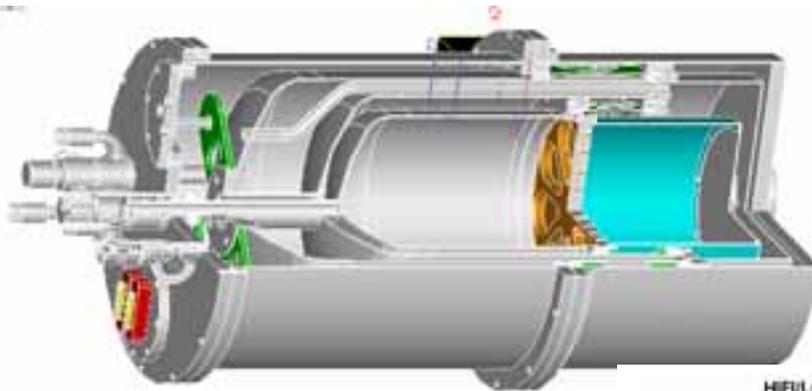

**Figure 13** (left) Dewar design for each of the GREAT frequency bands.

**Figure 14** (lower left) Mixer on SiN membrane mounted in flip-chip technique.

**Figure 15** (below) Out put power of the solid state LO amplifier chain at various frequencies. The horizontal line represents the goals and the solid curves the achievements.

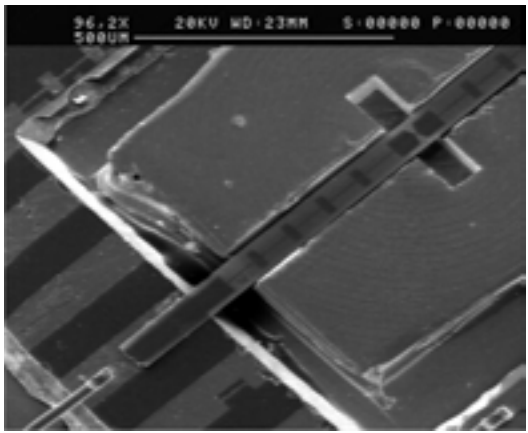

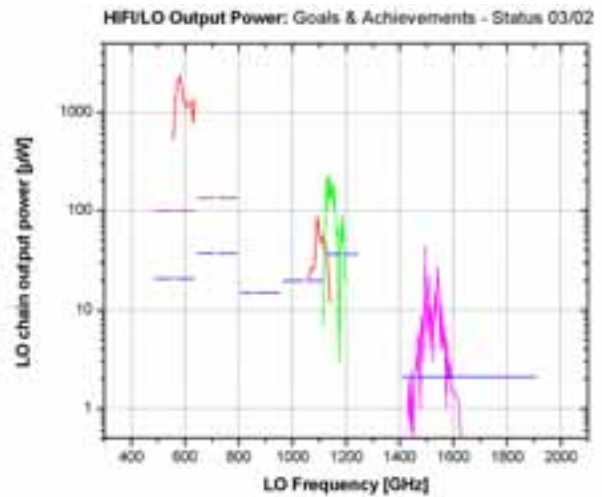

**CASIMIR**

CASIMIR is a multichannel heterodyne spectrometer with a spectral resolution of up to $10^6$ over the sub-millimeter and far-IR wavelengths range of 0.5 – 2 THz or 150 to 600 micron (Figure 17). A combination of advanced SIS and Hot Electron Bolometers (HEB) receivers will be used to cover this frequency range with seven bands. Tunnel-junction (SIS) mixers will cover the 0.5 – 1.2 THz range in four bands, while diffusion cooled HEBs will be used for three high frequency bands 1.2 – 2.0 THz. CASIMIR will use only solid state local oscillators (see above) with quasioptical coupling and mixers. Only the SIS mixers are expected at first light with CASIMIR.

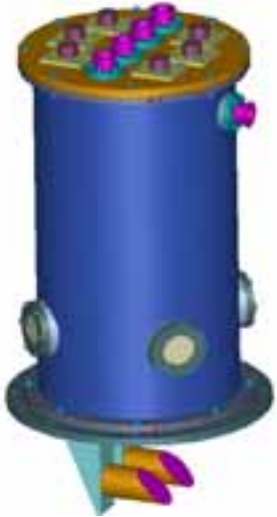
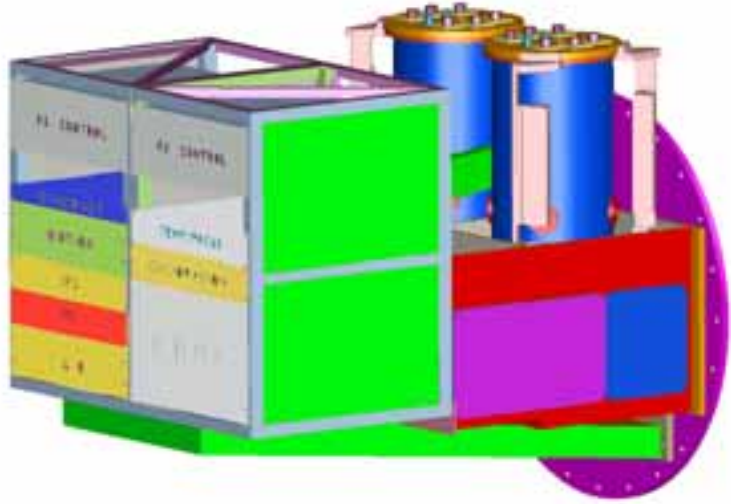

**Figure 16** Sketch of a CASIMIR cryostat. The SIS and HEB receivers will be housed in separate cryostats.

**Figure 17** Sketch of the CASIMIR instruments. Both CASIMIR as well as GREAT look very similar. Both are dual-channel systems with an optics box in common. The flange is on the right.

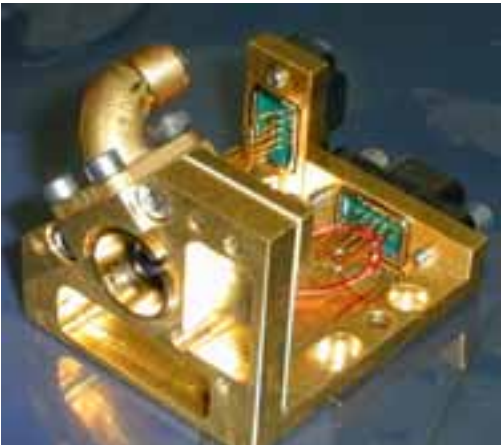
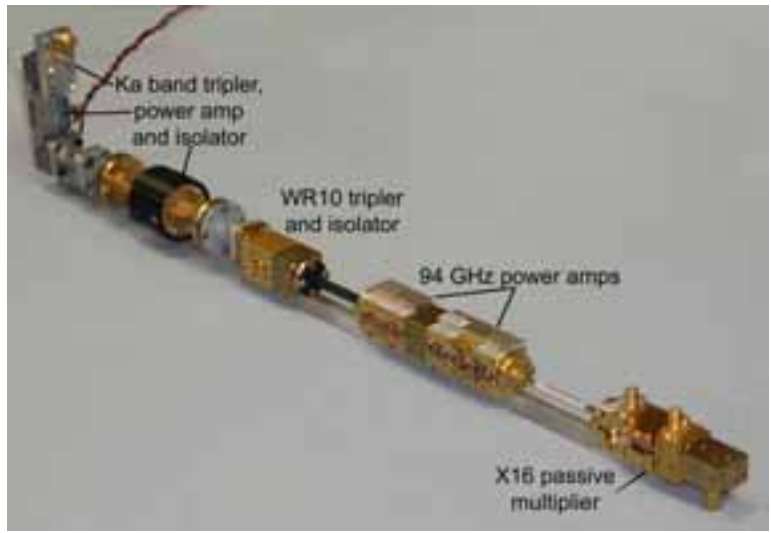

**Figure 18** New 1.2 THz SIS mixer.

**Figure 19** Solid state local oscillator providing several microwatts at 1.5THz.

The LO system consists of a computer controlled microwave frequency synthesizer. The microwave drives an active multiplier chain followed by a millimeter-wave HEMT power amplifier, the millimeter-wave source (MWS). There is a separate MWS for each mixer, so there will be up to 2 per cryostat (Figure 16), mounted on the sides of the cryostats as shown in Figure 17. The output from each MWS is coupled to a dedicated frequency multiplier (FM) via a wave-guide. These FMs will multiply the frequency of the MWS by a factor from 4 – 12, in order to provide the LO signal at 0.5 – 2 THz. An example of such an LO is shown in Figure 21, providing several microwatts at 1.5THz.

The development of the new 1.2 THz SIS mixer is proceeding well (Figure 19). New devices with smaller junction areas are available now. The power requirements for the LO have been reduced by using beam splitter injection. A 4 – 8 GHz cryogenic low noise amplifier has been designed and fabricated using Quasi-Monolithic Microwave Integrated circuit (QMMIC) technology (Figure 19). The QMMIC consists of a thin-film passive circuit on a GaAs substrate with three 160 micron gate InP HEMTs bump-bonded to it. At 4K, the amplifier drew a total of only 8 mW bias power and the noise temperature varied smoothly between 6K and 10 K. The 4 GHz bandwidth provided by this IF is necessary to allow observations of galaxies at the short sub-millimeter wavelengths. At the spatial resolution provided by SOFIA in the sub-millimeter wavelengths band, spectral features of galaxies are usually broadened due to the galactic rotation.

**EXES**

EXES is a 5 – 8 micron and 13 – 28 micron echelon cross echelle spectrograph operating in three spectroscopic modes: $\lambda/\Delta\lambda \approx 10^5$ (8 – 16 arcsec slit), $2 \times 10^4$, and 4000 (60 arcsec slit each). It uses an echelon, a 7.62 mm coarsely ruled, 93 degree steeply blazed diffraction grating of format 1m × 0.1 m to achieve high resolution (Figure 21). Cross dispersion is done with an echelle used at relatively low order. The detector is a 256 × 256 SiAs IBC array. A very similar instrument (TEXES) has already been deployed at the McDonald Observatory and the Infrared Telescope Facility (Figure 20).

EXES is well under construction: Detector and control electronics are all available and tested. The optical design is complete and the echelon is available. The mechanical design is complete and fabrication of the dewar will start by end of the year. Software and control is available and has been tested with TEXES on several previous observing runs.

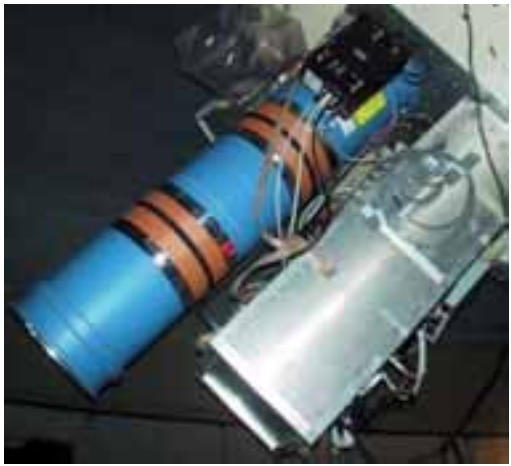
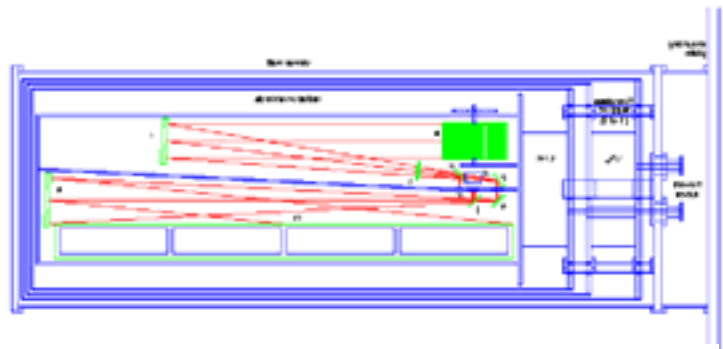

**Figure 20** (left) TEXES, the ground based predecessor of EXES mounted at a telescope.

**Figure 21** (top) Sketch of EXES. The echelon is the dominant component in the dewar.

**FIFI –LS**

FIFI-LS is a field imaging far IR line spectrometer covering two wavelengths bands, 42 – 110 micron and 110 – 220 micron, with two mid-resolution (R ≈ 2000) gratings operating simultaneously side by side (Figure 22). The design of the instrument is driven by the goal of maximizing observing efficiency, especially for observations of faint, extragalactic objects. Thus, FIFI LS utilizes an integral field technique that uses slicer mirrors to optically re-arrange the two-dimensional field into a single slit for a long slit spectrometer. Effectively, a 5 × 5 pixel spatial field of view is imaged to a 25 × 1 pixel slit and dispersed to a 25 × 16 pixel, two-dimensional detector array, providing diffraction-limited spatial and spectral multiplexing. Overall, for each of the 25 spatial pixels, the instrument can cover a velocity range of ≈ 1500 km/s around selected far-IR spectral lines with an estimated sensitivity of $2 \times 10^{-15}$ W Hz$^{1/2}$ per pixel. This arrangement provides good spectral coverage with high responsivity.

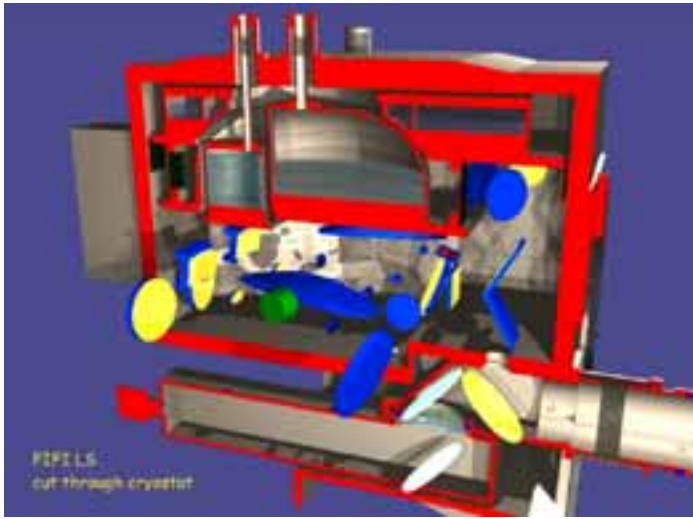

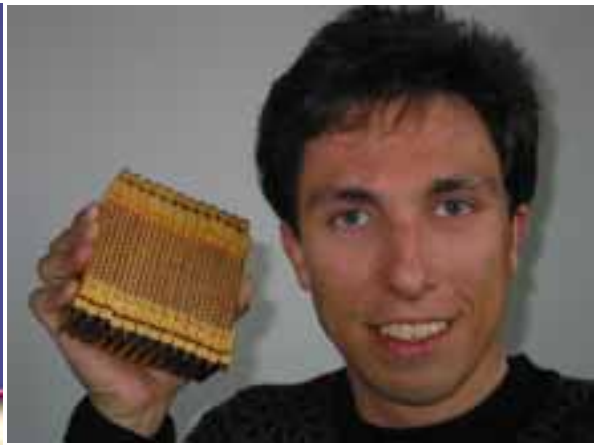

**Figure 22** (upper left) Sketch of the FIFI-LS experiment. Light enters from the right. The visible light is tracked by a camera in the lower section, the IR light is split off by a dichroic and fed into the cryostat in the upper section.

**Figure 23** (upper right) After months of hard work: Stressed Ge:Ga detector array fully assembled and contacted.

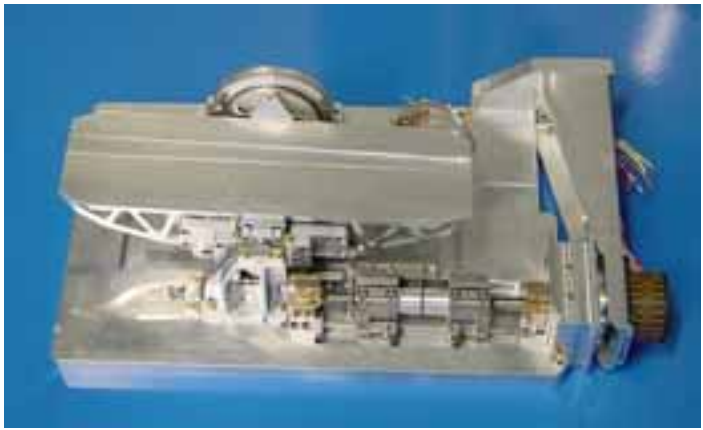

**Figure 24** (left) grating drive with stepper motor, gear and encoder. The length of the grating is 30 cm (1 foot).

Two sets of 16 × 25 pixel GeGa detectors, one stressed and one unstressed, have been developed for FIFI-LS. The design for the stressed and unstressed arrays is almost identical, except that for the unstressed array a residual stress will be applied for stability reasons. The stressed array has been fabricated, including two spares and is has successfully passed cryogenic testing (Figure 23). The quantum efficiency was measured 36% at 60mV bias.

The diffraction limited optics is mostly fabricated and will be completely delivered in August. The blue channel grating has been delivered; the red channel grating is in fabrication. The cryostat design is complete as is the design of the optomechanical components (Figure 24). The warm electronics is mostly designed. The first board of the cold-readout-electronics (CRE) is expected this summer.

**SAFIRE**

SAFIRE is a versatile imaging Fabry-Perot-interferometer (FPI) spectrograph covering 145 – 655 microns with spectral resolving powers ranging typically over 1000 – 10,000. A grating mode has also been included for long-slit spectroscopy over the same wavelength range with a resolving power of ≈ 100. The high resolution mode uses a tandem FPI with the low order FPI acting as the order sorter. For the lower spectral resolution only the low-order FPI will be used. Apart form the FPIs and the band-pass filters, the rest of the optics is all reflective.

The instrument uses two 16 × 32 filled arrays of superconducting transition edge sensor (TES) pop-up bolometers developed at NASA Goddard Space Flight Center (GSFC). The pop-up architecture (see HAWC above) allows for an ideal intimate connection of detectors and amplifiers in close-packed two-dimensional arrays, in the same spirit as the large 2d-arrays that revolutionized near- and mid-IR astronomy. The two separate detector technologies are required to provide adequate sensitivity over the full range of spectral resolutions, hence background powers. The detectors will be cooled to 0.065 K using an adiabatic demagnetization refrigerator (ADR, see HAWC).

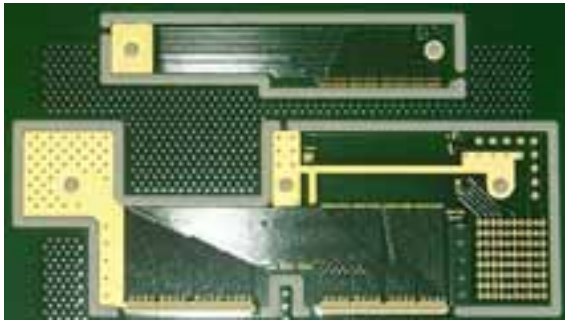
**Figure 25** Circuit board of SAFIREs pop-up detectors.

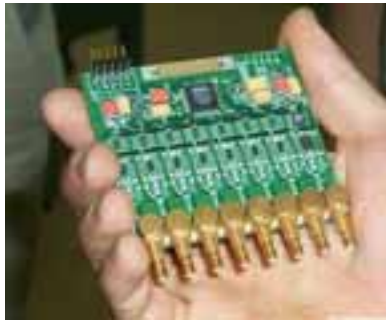
**Figure 26** Warm de-multiplexer electronics board.

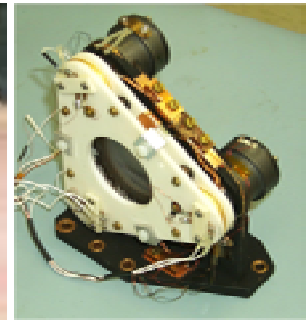
**Figure 27** Low-order Fabry-Perot-Interferometer, successfully tested at CSO last year.

The progress on this instrument has been somewhat slower compared to the other, mainly due to the challenging task of the detector development and due to the fact that SAFIRE is funded as a level of effort project. However, the test cryostat for SAFIRE is available, first batches of the bolometers have been tested and the design is making progress. Circuit boards for the cold and worm readout electronics have been developed (Figure 25 and 26) and are ready to use. The low-order FPI has been successfully tested at the CSO telescope (Figure 27). Motors with superconducting windings have been successfully tested and implemented.

## 3. SPECIAL INSTRUMENTS

**HIPO**

HIPO (formerly HOPI) is a special-purpose science instrument for SOFIA that is designed to provide simultaneous high-speed time resolved imaging photometry at two filter selectable optical wavelengths in the spectral range 0.35 – 1.1 micron (Figure 28). HOPI and FLITECAM will provide the option of being mounted on the SOFIA telescope simultaneously to allow data acquisition at two optical wavelengths and one near-IR wavelength (Figure 29). HOPI has a flexible optical system and numerous readout modes, allowing many specialized observations to be made. The instrument characteristics required for our proposed scientific pursuits are closely aligned to those needed for critical tests of the completed SOFIA Observatory.

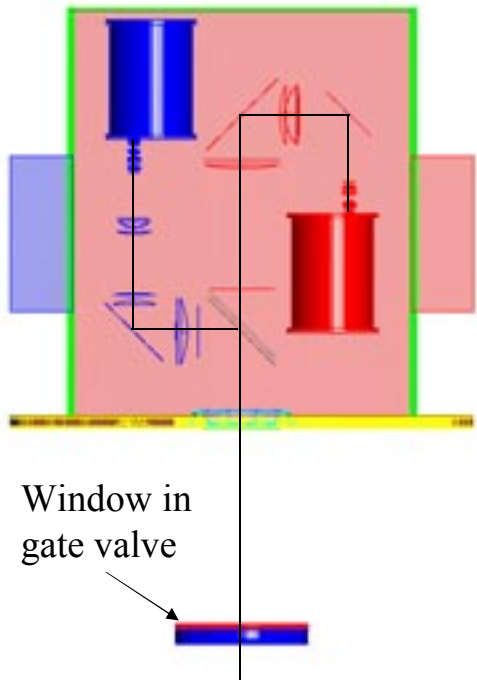
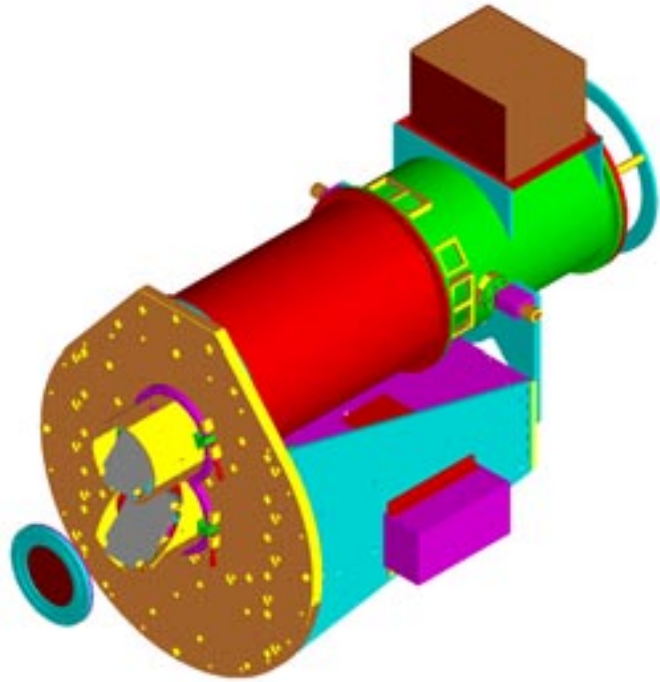

Window in gate valve

**Figure 28** Sketch of the HIPO instrument. The red (right) and blue channels (left) are separated by a beamsplitter.

**Figure 29** HIPO co-mounted below FLITECAM. A warm dichroic passes the optical beam into HIPO, while the IR light is reflected to FLITECAM.

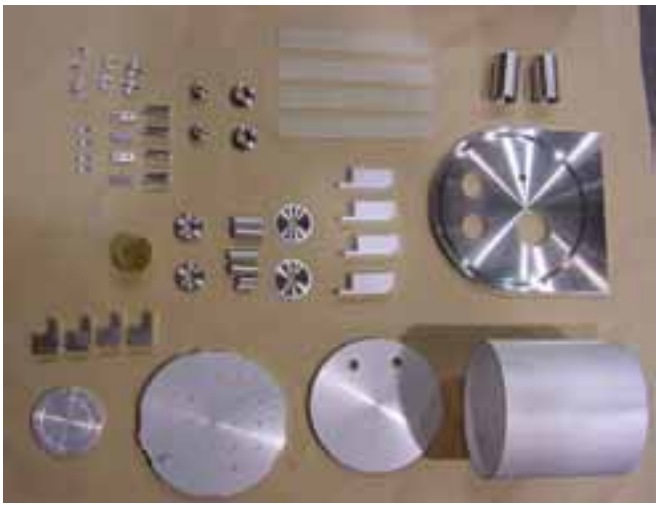

**Figure 30** Parts of the HIPO dewar, ready to be welded and integrated.

The primary scientific application HOPI is designed for is observation of stellar occultations by solar system objects. Occultations are fast events, the shadow velocity being comparable to the Earth's orbital velocity of -30 km/set. Integration times can therefore be as short as -10 ms. The Marconi EEV CCD47-20 detector with 13 micron pixels was selected for both HOPI channels. This device is thinned, backside illuminated, and anti-reflection coated for excellent quantum efficiency ($\geq$ 85% peak). The CCD47-20 is a frame transfer device and has a fast, high-gain on-chip amplifier. The read noise at slow scan speeds is -3 electrons rising to -6 electrons at 1 Mpx/sec read rate. Peak frame rate is 50 Hz.

The optics allows for various settings and scales including pupil viewing to be used as a Shack-Hartmann sensor. The seeing limited pixel scale is 0.33 arcsec/pixel. The instrument is cooled with liquid nitrogen.

The cryostat design has been approved and fabrication has started (Figure 30). Welding is scheduled for August 02. All lenses are in house. All other optics components are being fabricated. Flight version of GPS (Global Positioning System) delivered. Prototype timing circuits are being tested. The CCD electronics is in house.

## 4. SUMMARY

SOFIA is the telescope with the widest spectral range ever build (0.3 – 1500 micron). Consequently, the SOFIA first light instruments span an impressive range from the ultraviolet to the sub-millimeter and present a diversity in instrument concepts not found at any other observatory. There is good reason to believe that all first light instruments will be ready when SOFIA takes off for the first time.

The first call for observing proposals will probably be released in fall of 2003. An update of the status of the call for proposals can be found on the web page: http://sofia.arc.nasa.gov/Science/sci.html.